%% file: main.tex
\documentclass[12pt]{article}

\usepackage{enumitem,amssymb}
\usepackage{ragged2e}
\usepackage{graphicx}
\usepackage{comment}
\usepackage[utf8]{inputenc}
\usepackage{charter}
\usepackage{fullpage}
\usepackage{hyperref}
\usepackage{graphicx}
\usepackage{lipsum}
\usepackage[sort&compress,numbers]{natbib}
\usepackage{hyperref}
\hypersetup{hidelinks}
\usepackage[total={6.5in,9in},top=1in,headsep=0.1in,headheight=1in]{geometry}
\usepackage[utf8]{inputenc}

\usepackage{aas_macros}

\usepackage{times}
\usepackage{geometry}
\usepackage{enumitem,amssymb}
\usepackage{comment}
\usepackage{multicol}
\usepackage[caption=false]{subfig}
\usepackage[usenames]{xcolor} 
\definecolor{xlinkcolor}{cmyk}{1,1,0,0}
\usepackage{enumitem}
\setenumerate{itemsep=0mm}

\newcommand\snowmass{
\begin{center}
  \rule[-0.2in]{\hsize}{0.01in}\\
  \rule{\hsize}{0.01in}\\
  \vskip 0.1in
Submitted to the Proceedings of the US Community Study\\
  on the Future of Particle Physics (Snowmass 2021)\\
  \rule{\hsize}{0.01in}\\
  \rule[+0.2in]{\hsize}{0.01in}\\[-2em]
\end{center}
}

\usepackage[firstpage=true]{background}
\backgroundsetup{contents={\parbox{6.5in}{\snowmass}}, scale=1,placement=top,opacity=1,color=black,position={3.25in,1.2in}}

\usepackage{fancyhdr}
\fancypagestyle{plain}{%
  \fancyhf{}%
  \fancyhead[C]{}
  \fancyfoot[C]{\thepage}
}

\fancypagestyle{empty}{%
  \fancyhf{}%
  \fancyhead[C]{{\it Snowmass2021 Cosmic Frontier White Paper Template}}
  \fancyfoot[C]{\thepage}
}
\title{Snowmass2021 Cosmic Frontier White Paper:
\\Numerical relativity for next-generation gravitational-wave probes of fundamental physics}
\date{}

\usepackage{authblk}

\author[1]{Francois Foucart}
\author[2]{Pablo Laguna}
\author[3]{Geoffrey Lovelace}
\author[4]{David Radice}
\author[5]{Helvi Witek}

\affil[1]{Department of Physics \& Astronomy, University of New Hampshire, Durham, New Hampshire 03824, USA}
\affil[2]{Center for Gravitational Physics and Department of Physics, The University of Texas at Austin, Austin, TX 78712, USA}
\affil[3]{Nicholas and Lee Begovich Center for Gravitational-Wave Physics and Astronomy, California State University, Fullerton, Fullerton, CA 92834, USA}
\affil[4]{Institute for Gravitation \& the Cosmos, The Pennsylvania State University, University Park PA 16802, USA}
\affil[5]{Illinois  Center  for  Advanced  Studies  of  the  Universe and Department of Physics, University of Illinois at Urbana-Champaign, Urbana, Illinois 61801, USA}

\begin{document}

\maketitle

\begin{abstract}
The next generation of gravitational-wave detectors, conceived to begin operations in the 2030s, will probe fundamental physics with exquisite sensitivity. These observations will measure the equation of state of dense nuclear matter in the most extreme environments in the universe, reveal with exquisite fidelity the nonlinear dynamics of warped spacetime, put general relativity to the strictest test, and perhaps use black holes as cosmic particle detectors. Achieving each of these goals will require a new generation of numerical relativity simulations that will run at scale on the supercomputers of the 2030s to achieve the necessary accuracy, which far exceeds the capabilities of numerical relativity and high-performance computing infrastructures
available today.
\end{abstract}

\tableofcontents


\section{Motivation}\label{sec:motivation} Gravitational waves are ripples of
warped spacetime that travel at the speed of light. In 2015, a century after
Einstein predicted their existence, the Laser Interferometer Gravitational-Wave
Observatory (Advanced LIGO) discovered gravitational waves from a merging binary
black hole as the waves passed through Earth~\cite{LIGOScientific:2016aoc}. Two
years later, LIGO and Virgo observed gravitational waves from merging neutron
stars~\cite{TheLIGOScientific:2017qsa}, a collision also observed by telescopes
spanning the electromagnetic spectrum~\cite{LIGOScientific:2017ync}. These
events, together with the dozens of gravitational waves that LIGO and Virgo have
observed, have inaugurated the era of gravitational-wave
astronomy~\cite{LIGOScientific:2020ibl, LIGOScientific:2018mvr,
LIGOScientific:2021djp}.

The next generation of gravitational-wave detectors will use gravitational waves
from sources throughout the cosmos to probe fundamental physics with
unprecedented sensitivity, as discussed in a separate Snowmass White
Paper~\cite{Berti:2022wzk}. Proposed detectors on Earth include LIGO
Voyager~\cite{Adhikari:2019zpy}, Cosmic Explorer~\cite{Reitze:2019iox}, Einstein
Telescope~\cite{Punturo:2010zz}, and NEMO~\cite{Ackley:2020atn}; future
ground-based gravitational-wave facilities are described in a separate Snowmass
~\cite{BallmerSnowmass2022}.
In space, the Laser Interferometer Space Antenna (LISA)~\cite{LISA:2017pwj}, the
DECi-hertz Interferometer Gravitational-wave Observatory
(DECIGO)~\cite{Kawamura:2020pcg}, and TianQin~\cite{TianQin:2015yph} will
observe gravitational waves at frequencies too low to ever detect on Earth
because they would be obscured by seismic noise.

Next-generation gravitational-wave detectors
are anticipated to begin observations in the 2030s. Their observations of
coalescing binary neutron stars and black-hole/neutron-star binaries will
measure the equation of state of dense nuclear matter in the most extreme
environments in the universe, and their observations of gravitational waves from
merging black holes---which contain the strongest spacetime curvature in the
universe---will put general relativity to the strictest tests. These future
gravitational-wave detectors might also enable observations that use
black holes as cosmic particle detectors, potentially giving new, complementary
insight into the nature of dark matter.

Accurate theoretical models of gravitational waves are critical for interpreting
gravitational-wave observations---specifically, for inferring the nature and behavior of their
sources. Long before the time of coalescence, the gravitational waves from
merging black holes and neutron stars can be well modeled using the
post-Newtonian approximation, which approximates general relativity in the limit
of weak gravity and small velocities. Long after the time of coalescence, the
gravitational waves from a black hole remnant resulting from merging black holes
and neutron stars can be well approximated using perturbation theory. But near
the time of coalescence, when the spacetime curvature and (if present)
neutron-star matter are the most nonlinear and dynamic, all known analytic
approximations break down: the emitted gravitational waves and the
strong-gravity dynamics of their source can only be calculated with numerical
relativity.

Numerical relativity amounts to numerically solving the equations of general
relativity or, for simulations involving neutron stars, the equations of general
relativistic radiation magnetohydrodynamics (fluid dynamics, magnetic fields, and radiation transport); the techniques of numerical relativity are
reviewed, e.g., in Refs.~\cite{Faber:2012rw, Baiotti:2016qnr, Dietrich:2018phi,
Duez:2018jaf,Dietrich:2020eud}, and briefly discussed in Sec.~\ref{sec:methods}.
Numerical-relativity calculations are technically challenging, in part because
the equations are strongly nonlinear and,
in the presence of neutron-star matter,
because the solutions contain small scale features that are especially challenging to resolve, such as shocks, neutron-star surfaces, and turbulence. These calculations are
also computationally expensive, requiring high-performance computing to achieve
the necessary accuracy.

\begin{figure}
  \centering
  \includegraphics[width=5in]{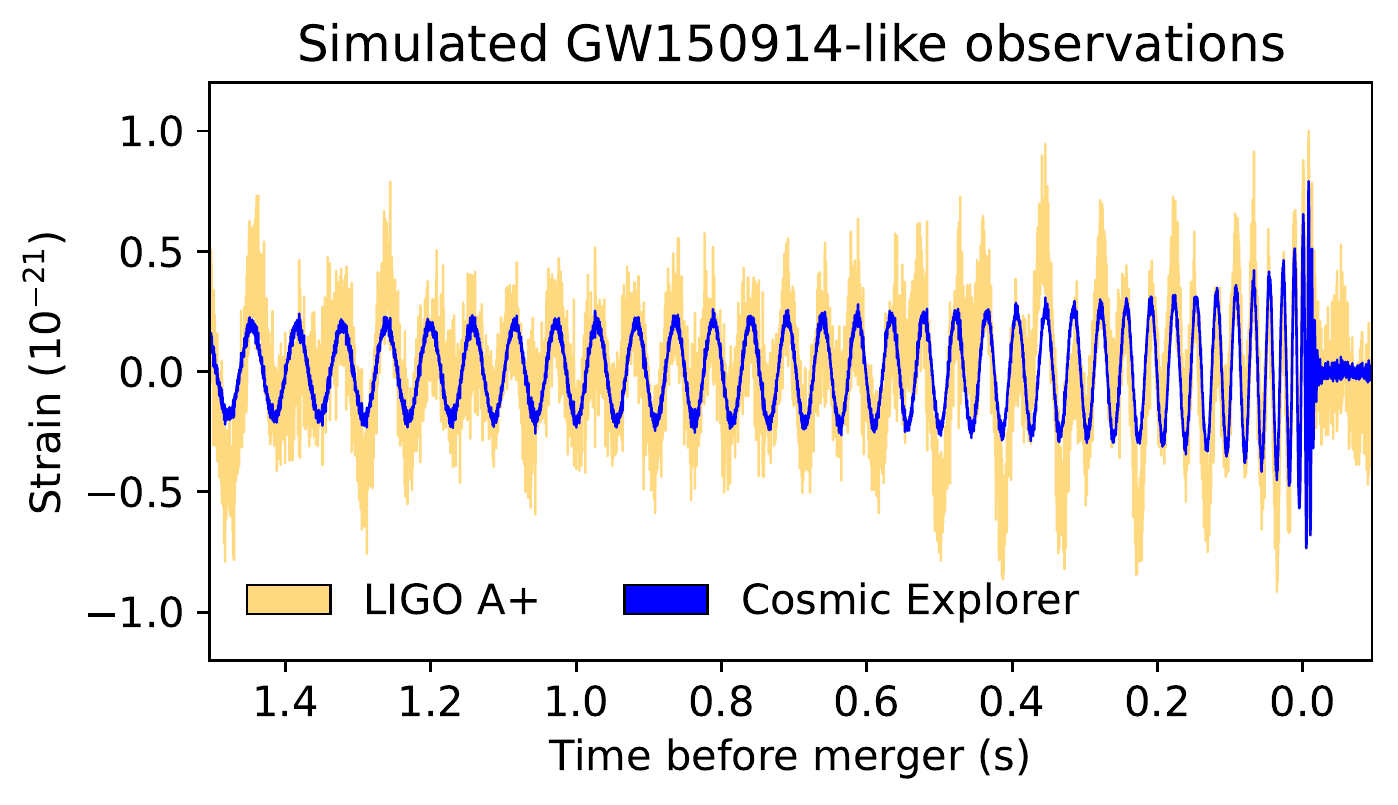}
\caption{Simulated gravitational-wave detector strain measurements of
  gravitational waves from two merging black holes. The signal is similar to
  GW150914~\cite{LIGOScientific:2016aoc}, the first directly detected
  gravitational waves. The strain is shown as a function of time for the signal
  superimposed on both simulated Cosmic Explorer noise (blue) and simulated LIGO
  A+ noise (yellow). Taken from Fig.~5.2 of Ref.~\cite{Evans:2021gyd}.
  \label{fig:GW150914vsNoise}}
\end{figure}

How much accuracy is enough?	The answer depends on the signal-to-noise ratio of
an observation: roughly speaking, avoiding any bias in gravitational-wave
interpretation requires numerical uncertainties smaller than the observation's
measurement uncertainty. The observations with the most potential to reveal new
fundamental physics are those with the highest signal-to-noise
ratios---precisely those observations that demand the most accuracy from
numerical-relativity models. And the observations with the highest
signal-to-noise ratios will come from next-generation detectors: their loudest
observations will have signal-to-noise ratios in the thousands, more than an
order of magnitude beyond the strongest signals observed to date.
Figure~\ref{fig:GW150914vsNoise} illustrates this gain in sensitivity by showing
two simulated gravitational-wave detections of the same gravitational-wave
source, one using an upgraded LIGO detector, and the other using Cosmic
Explorer, a next-generation detector.

Extracting information from such high-fidelity signals while limiting systematic
biases will require models with an order of magnitude increase in accuracy over
today's state of the art. Achieving this accuracy will require a new generation
of numerical-relativity software, designed to run at scale on the exascale
supercomputers that will be available in the 2030s. New (typically open-source)
numerical-relativity codes under development today will help meet this goal by
producing publicly available catalogs of simulated gravitational waveforms for
coalescing compact binaries (i.e., binary black holes, binary neutron stars, and
black-hole/neutron-star binaries).

The rest of this whitepaper is organized as follows. Sec.~\ref{sec:methods}
briefly summarizes the methods of
gravitational waveform modeling,
especially numerical relativity. Then, Sec.~\ref{sec:ns} discusses progress and
challenges in applying numerical relativity to probe nuclear physics and the
nature of neutron stars. Sec.~\ref{sec:highprecision} explains in more
quantitative detail the challenge that high-precision gravitational-wave
observations pose to numerical-relativity waveform modeling. In
Sec.~\ref{sec:testinggr}, we discuss the importance of numerical-relativity
waveform modeling in using gravitational-wave observations to seek physics
beyond general relativity, and in Sec.~\ref{sec:particle} we discuss numerical
relativity's role in the possibility of using black holes as cosmic particle
detectors. Finally, in Sec.~\ref{sec:conclusion} we present a brief summary and
discuss the future work needed to fully realize the potential of gravitational
waves as probes of fundamental physics.



\section{Gravitational waveform modeling}\label{sec:methods}
\begin{figure}
  \centering
  \includegraphics[width=5in]{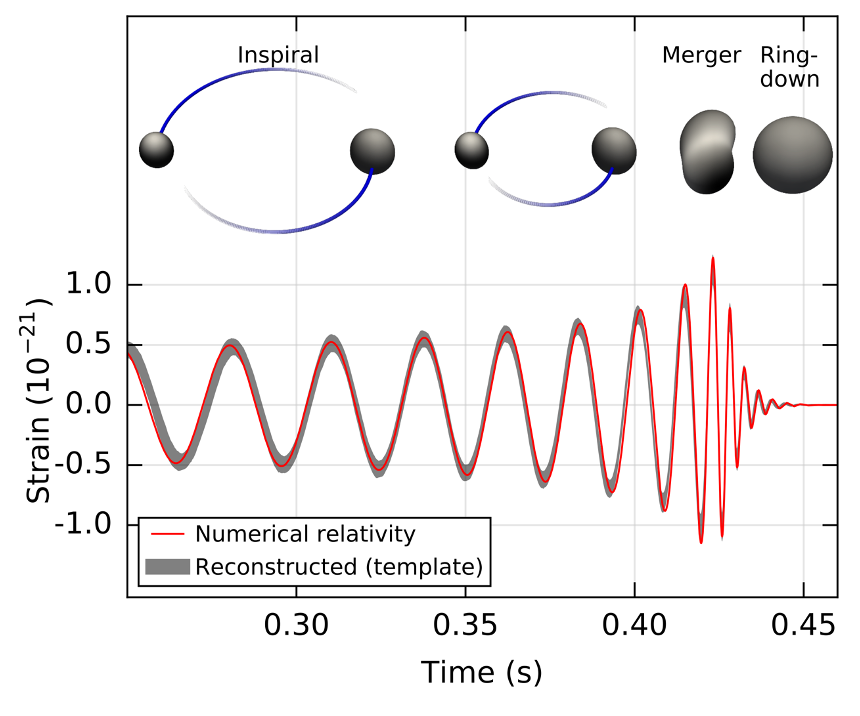}
  \caption{\label{fig:GW150914NRWaveform}
    Numerical relativity waveform modeling GW150914, the first gravitational wave signal detected by LIGO.
    The inset shows the black holes' horizons during the inspiral, merger and ringdown.
    Taken from Fig.~2 of Ref.~\cite{LIGOScientific:2016aoc}.
  }
\end{figure}

A gravitational wave signal encodes vital information about its sources, such as
the masses and spins of the companions in a compact binary, the equation of
state of dense matter if one of them is a neutron star, and the underlying
theory of gravity.
These parameters are identified by using matched filtering techniques, in which
the observed signal is compared against a catalog of theoretical gravitational
waveform models, called templates.
The templates need to accurately cover the different phases of a binary's
evolution consisting of the inspiral, merger and ringdown
illustrated in Fig.~\ref{fig:GW150914NRWaveform}.
%

Models of a binary's evolution typically rely on two core methods:
(i) approximations, such as Post-Newtonian (PN) or Post-Minkowskian
expansions, that are suitable for modeling the early inspiral of a compact
binary using a weak-field and small velocity expansion; and (ii) numerical
relativity, which numerically calculates a binary's late inspiral, merger and
ringdown by solving Einstein's equations (or extensions of them) in the nonlinear regime.
Both core methods feed into the production of full inspiral-merger-ringdown templates using either
phenomenological models that directly combine PN
and numerical relativity waveforms (e.g.~\cite{Pratten:2020ceb}),
effective-one-body models (e.g.~\cite{Ossokine:2020kjp}) that are a resummation
of the PN expansion and are calibrated against numerical relativity, or
surrogate models (e.g.~\cite{Varma2019_NRSur7dq4}) that directly interpolate
numerical-relativity waveforms. To ensure that the gravitational-wave
interpretation is not limited by modeling errors, even as future gravitational
wave detectors achieve ten to hundred times better sensitivity than today's
detectors, highly accurate waveform templates are crucial.
In this white paper, new advances and challenges in numerical relativity are discussed, while new developments in using scattering amplitudes and effective field theory for gravitational-wave modeling are presented in a separate Snowmass White Paper~\cite{BuonannoSnowmass}.
A more extensive review on waveform modeling for future gravitational wave detectors can be found in the LISA Waveform Working Group White Paper~\cite{LISAWavWGWP}.

Before outlining new physical applications and computational challenges below, we here give a
brief summary of the current status of numerical relativity.
Numerical relativity refers to solving Einstein's equations, or extensions of them,
possibly coupled to matter or additional fundamental fields,
in four spacetime dimensions.
This typically requires high-performance computing, because the equations form a system of more than ten coupled,
nonlinear, partial differential equations (PDEs) of mixed character.
By applying a spacetime decomposition into three dimensional, spatial hypersurfaces that are then propagated
in time,
the equations can be formulated as a time-evolution problem, subject to a set of constraints.
Using this approach, a numerical-relativity calculation is  divided into three stages:
\begin{enumerate}
\item Construction of initial data that represents the initial configuration (e.g., two compact objects orbiting each other in equilibrium). This requires solving the constraint
equations, a set of coupled, elliptic-type PDEs in three dimensions. The bulk of contemporary numerical relativity software uses either the Bowen-York conformal approach
or the conformal thin-sandwich method.
\item Time evolution, i.e., a binary's development in time that is encoded in a set of coupled, hyperbolic-type PDEs
and must be complemented by suitable gauge conditions.
The majority of the numerical relativity codes uses either
a variant of the generalized harmonic formulation of Einstein's equations together with the damped harmonic gauge~\cite{Pretorius2005a,Lindblom:2005qh,Szilagyi:2009qz}
or a variant of the Z4~\cite{Bona:2003fj,Bernuzzi:2009ex,Weyhausen:2011cg,Alic:2011gg,Alic:2013xsa} or Baumgarte-Shapiro-Shibata-Nakumara (BSSN) formulations~\cite{Baumgarte99,Shibata1995}
that are complemented by the moving puncture approach~\cite{Campanelli:2005dd,Baker:2005vv}.
\item Extraction of physical information such as the gravitational and additional radiation or, in case of black hole spacetimes, the apparent horizons. Note that for modeling observations in distant gravitational-wave detectors, the gravitational radiation must be propagated to future null infinity (see Ref.~\cite{Bishop:2016lgv} for a review), either by extrapolation~\cite{Boyle:2009vi}, by evolving it, e.g. Cauchy-Characteristic Evolution (CCE)~\cite{Bishop:1996gt}, or through perturbative techniques~\cite{Nakano:2015pta}, and gauge conditions and transformations at future null infinity must be treated with care to yield well-behaved numerical waveforms.
\end{enumerate}

\begin{table}[!htbp]
\centering
\begin{tabular}{l l l l l l} \hline
Code & Open Source & Catalog & Formulation & Hydro & Beyond GR  \\ \hline
AMSS-NCKU~\cite{PhysRevD.78.124011,PhysRevD.87.104029,PhysRevD.85.124032,PhysRevD.88.084057} & Yes & No & BSSN/Z4c & No & Yes\\
BAM~\cite{Bruegmann:2006ulg,Thierfelder:2011yi,Dietrich:2015iva} & No & \cite{Dietrich:2018phi} & BSSN/Z4c & Yes & No \\
BAMPS~\cite{Bugner:2015gqa,Hilditch:2015aba} & No & No & GHG & Yes & No \\
COFFEE\cite{Doulis:2019ejj,Frauendiener:2021eyv} & Yes &  No & GCFE & No & Yes \\
Dendro-GR~\cite{Fernando:2018mov,fernando2019massively,fernando2019scalable} & Yes & No & BSSN/CCZ4 & No & Yes\\
Einstein Toolkit~\cite{Loffler:2011ay,einsteintoolkit} & Yes &  No & BSSN/Z4c & Yes & Yes \\
$^{\ast}$Canuda~\cite{Canuda,Okawa:2014nda,Zilhao:2015tya,Witek:2018dmd} & Yes   & No    & BSSN  & No    & Yes \\
$^{\ast}$IllinoisGRMHD~\cite{Etienne:2015cea} & Yes & No & BSSN & Yes & No \\
$^{\ast}$LazEv~\cite{Campanelli:2005dd,Zlochower:2005bj} & No & \cite{Healy:2017psd,Healy:2019jyf,Healy:2020vre,Healy:2022wdn} & BSSN+CCZ4 & No & No \\
$^{\ast}$Lean~\cite{Sperhake:2006cy,Berti:2013gfa} & Partially & No & BSSN & No & Yes\\
$^{\ast}$MAYA~\cite{Jani:2016wkt} & No & \cite{Jani:2016wkt} & BSSN & No & Yes \\
$^{\ast}$NRPy+~\cite{Ruchlin:2017com} & Yes &  No & BSSN & Yes & No \\
$^{\ast}$SphericalNR~\cite{Mewes:2018szi,Mewes:2020vic}
& No & No & spherical BSSN & Yes & No \\
$^{\ast}$THC~\cite{Radice:2012cu, Radice:2013hxh, Radice:2013xpa} & Yes & \cite{Dietrich:2018phi} & BSSN/Z4c & Yes & No \\
ExaHyPE~\cite{Koppel:2017kiz} & Yes & No & CCZ4 & Yes & No \\
FIL\cite{Most:2019kfe} & No &  No & BSSN/Z4c/CCZ4 & Yes & No \\
FUKA~\cite{Papenfort:2021hod, FUKA} & Yes & No & XCTS & Yes & No\\
GR-Athena++~\cite{Daszuta:2021ecf} & Yes  &  No & Z4c & Yes & No\\
GRChombo~\cite{Clough:2015sqa, GRChombo,Andrade:2021rbd} & Yes & No & BSSN+CCZ4 & No & Yes \\
HAD~\cite{Had,Liebling:2002qp,Lehner:2005vc} & No & No & CCZ4 & Yes & Yes\\
Illinois GRMHD~\cite{Etienne:2010ui,Sun:2022vri} & No & Yes & BSSN & Yes & No \\
MANGA/NRPy+~\cite{Chang:2020ktl} & Partially &  No & BSSN & Yes & No \\
MHDuet~\cite{Palenzuela:2018sly,Liebling:2020jlq}    &No           &          No   &                    CCZ4         & Yes    &       Yes \\
SACRA-MPI~\cite{Kiuchi:2019kzt} & No & & BSSN+Z4c & Yes & No \\
SpEC~\cite{SpECWebsite,Boyle:2019kee}         & No  & \cite{Boyle:2019kee,SXS:catalog}  & GHG & Yes & Yes \\
SpECTRE~\cite{Kidder:2016hev,spectre} & Yes &   No      & GHG & Yes & No    \\
SPHINCS\_BSSN~\cite{Rosswog:2020kwm} &  & No & BSSN & SPH & No \\

\hline
\end{tabular}
\caption{\label{tab:NRcodesLISAWavWG}
List of numerical relativity codes.
We indicate if a code is open-source, if it has been used to produce gravitational waveform catalogs, the formulation of Einstein's equation used
(GHG: generalized harmonic,
BSSN: Baumgarte-Shapiro-Shibata-Nakamura,
CCZ4 / Z4c variants of the Z4 formulation,
GCFE: generalised conformal field equations
), if a code implements general relativistic hydrodynamics, and if it is capable to simulate compact binaries beyond general relativity. An asterisk indicates codes that are either (partially) based on the open-source Einstein Toolkit or are co-funded by its grant.
Credit: Deidre~Shoemaker; taken from Ref.~\cite{LISAWavWGWP}.
}
\end{table}

Since the breakthroughs in numerical relativity in 2005~\cite{Pretorius2005a}
and 2006~\cite{Campanelli:2005dd,Baker:2005vv}, that saw the very first simulations of the last orbits of a black hole binary and its merger,
the field has matured into a state-of-the-art tool
to investigate extreme gravity.
A large variety of numerical relativity cyberinfrastructures for computational astrophysics is available.
In Table~\ref{tab:NRcodesLISAWavWG}, we present a list of currently available numerical-relativity software, indicating for each if it is open-source, the formulation of Einstein's equations used, if it is capable of performing general relativistic hydrodynamics simulations (and not just vacuum simulations), and if it is capable to perform simulations in alternative theories of gravity. This list is adapted from the LISA Waveform Working Group White Paper~\cite{LISAWavWGWP}.

A number of the numerical relativity codes (collaborations) have constructed
catalogs of simulated gravitational waveforms as indicated in Table~\ref{tab:NRcodesLISAWavWG}.
For black-hole binaries, the combined catalogs contain more than $5,700$
waveforms that cover mass ratios $q=m_{1}/m_{2}=1,\ldots,15$ up to $q=128$,
where $m_{1}$ ($m_{2}$) is the mass of the heavier (lighter) black hole, and
spins magnitudes up to $0.998$
~\cite{Mroue:2013xna,Boyle:2019kee,Jani:2016wkt,Healy:2017psd,Healy:2019jyf,Healy:2020vre,Healy:2022wdn,
  Lousto:2020tnb}. There are also the first numerical relativity waveform catalogs
for binary neutron stars~\cite{Dietrich:2018phi}, and a recent
study~\cite{Lousto:2022hoq} used head-on collisions (in which the black holes
begin at rest) to demonstrate that numerical-relativity techniques can in
principle model gravitational-wave emission at mass ratios as high as 1000.

Given the wealth of available simulations, what future development is needed?
The answer is two-fold and concerns the waveform accuracy as well as the physics included in the models. Each of these items will be discussed in detail in the following sections.


\section{Nuclear physics and neutron stars}\label{sec:ns}
When two neutron stars, or a black hole and a neutron star, coalesce, they emit gravitational waves that encode the behavior of the densest matter in the universe. The cold cores of neutron stars are expected to have densities $\rho \sim 10^{15}\,{\rm g/cm^3}$. In that regime, the strength of nuclear interactions between densely packed particles is uncertain, and even the composition of the core is unknown. The properties of dense matter are however tightly correlated with the size of neutron stars, their maximum mass, and their response to external gravitational fields. Dense matter's presence in a merging binary, as a finite size object distorted by the gravitational field of its companion, leads to more gravitational wave emission than for black hole binaries and a faster evolution towards merger \cite{Damour:2012yf, Read:2013zra, DelPozzo:2013ala, Bernuzzi:2014owa, Hotokezaka:2016bzh, Hinderer:2016eia, De:2018uhw, LIGOScientific:2018cki}. The size of a neutron star also determines if and when it can be tidally disrupted by a black hole companion (for black hole-neutron star binaries) and when two neutron stars collide and merge (for neutron star-neutron star binaries) \cite{Foucart:2012nc, Foucart:2012vn, Kyutoku:2015gda, Kyutoku:2021icp}. Finally, the post-merger evolution of a neutron star-neutron star binary is strongly impacted by the properties of dense matter: unknown nuclear physics determines whether the remnant collapses to a black hole, as well as the frequency of post-merger gravitational waves driven by oscillations in the remnant~\cite{Sekiguchi:2011mc, Hotokezaka:2011dh, Bauswein:2012ya, Takami:2014zpa, Radice:2016rys, Most:2018eaw, Bauswein:2018bma, Liebling:2020dhf, Bauswein:2020xlt, Radice:2020ddv, Most:2021ktk, Prakash:2021wpz, Perego:2021mkd, Breschi:2021xrx, Huang:2022mqp, Most:2022wgo, Kedia:2022nns}. Recovering this information from gravitational-wave observations requires an accurate theoretical understanding of the emitted waves and thus high-accuracy numerical relativity simulations.

These simulations are challenging and expensive, yet they must be sufficiently accurate to avoid introducing systematic biases into the interpretation of gravitational-wave observations. The accuracy required (cf.~Sec.~\ref{sec:highprecision}) increases with the square of the observation's signal-to-noise ratio~\cite{Lindblom:2008cm}. The first (and loudest) gravitational wave observation from coalescing binary neutron stars to date, GW170817~\cite{TheLIGOScientific:2017qsa}, had a signal-to-noise ratio (SNR) $\sim 30$. Recent studies~\cite{Samajdar:2018dcx, Gamba:2020wgg, Huang:2020pba} find that systematic uncertainties from inaccurate waveform models would be substantial at SNRs $\gtrsim 70$, which could be achieved if a signal as loud as GW170817 were observed in current-generation detectors when they achieve their design sensitivities. The tremendous sensitivity gains that future gravitational-wave detector concepts~\cite{Punturo:2010zz, Reitze:2019iox, Adhikari:2019zpy} would achieve means that they would observe a GW170817-like signal with an SNR in the thousands~\cite{Maggiore:2019uih}, requiring vastly more accurate theoretical waveform models.

Simulations modeling the tidal response of the neutron stars during the last
stages of the inspiral need to decrease their phase errors by more than two
orders of magnitude. With current simulation technology, this is expected to
require the grid resolution to be decreased by at least a factor 10 compared to
the highest resolution simulations available to date \cite{Dietrich:2017aum,
  Kiuchi:2017pte} (assuming second order convergence), leading to a $\sim 10^4$
increase in computational cost. Even if numerical relativity codes could scale
efficiently to millions of CPU cores, a single simulation would still require
several years to complete and tens of billions of CPU hours.

Neutron star mergers also power bright electromagnetic counterparts that carry additional information about the merging objects, including the properties of dense matter. Neutron-rich matter ejected during and after merger undergoes r-process nucleosynthesis, making neutron star mergers one of the lead candidates for the production site of r-process elements~\cite{1976ApJ...210..549L}. The radioactive decay of the ashes of the r-process powers {\it kilonovae}, UV/optical/infrared transients observable days to weeks after the merger~\cite{Li:1998bw,Roberts2011,2013ApJ...775...18B}. Some neutron star mergers also result in the formation of massive accretion disks around a compact object that power narrow jets of highly-relativistic material observed from Earth as {\it short gamma-ray bursts}~\cite{eichler:89,1992ApJ...395L..83N,2007NJPh....9...17L}. Both types of signals were observed following the first neutron star merger detection (GW170817)~\cite{2017ApJ...848L..13A}. Numerical simulations are required to understand which mergers power electromagnetic signals and to connect the observable properties of these signals to the properties of the merging compact objects and to the equation of state of dense matter.

Simulations aiming to model the post-merger gravitational wave signal of neutron star binaries and to study their electromagnetic counterparts face the additional challenge of having to resolve high Reynolds number magneto-hydrodynamics turbulence and to model complex neutrino-radiation effects, which might impact the postmerger gravitational wave signal and will certainly impact the electromagnetic counterparts and nucleosynthesis yields of these events \cite{Radice:2020ddv}. Extremely strong magnetic fields ($\sim 10^{16}\,{\rm G}$) are likely grown from small scale magneto-hydrodynamics instabilities in neutron star mergers, and even the highest resolution simulations performed to date (with grid spacing an order of magnitude smaller than what is typically affordable with current codes) have not been able to converge to a well-defined answer for the post-merger magnetic field~\cite{Kiuchi2015}. As magnetic fields are likely a crucial ingredient in the production of short gamma-ray bursts~\cite{Rezzolla:2011da,Paschalidis2014,Christie:2019lim} and in the ejection of the material producing r-process elements and kilonovae~\cite{Siegel:2017nub,Christie:2019lim}, this represents a major limitation in our ability to model these systems. Neutrino-matter interactions are less important to the dynamics of the post-merger remnant, but they play a major role in setting the composition (neutron-richness) of matter outflows, which largely determines the outcome of r-process nucleosynthesis~\cite{Wanajo2014}. Properly including all relevant neutrino processes is a daunting challenge. At the very least, we will need to solve the 7-dimensional transport equations; but even that may not be enough. For example, neutrino oscillations due to fast-flavor instabilities may significantly impact the composition of matter outflows~\cite{Li:2021vqj} and can only be captured by evolving the quantum kinetics equations with grid resolution orders of magnitude smaller than what is used in merger simulations.

	A direct approach using current codes and numerical methods cannot be successful. Instead, the numerical relativity community will need to develop more accurate numerical schemes to model tidally interacting neutron stars, sophisticated algorithms for neutrino-radiation hydrodynamics, and subgrid turbulence models. First steps in these directions have been made \cite{Radice:2013hxh, Fambri:2018udk, Kidder:2016hev, Foucart:2021mcb, Palenzuela:2021gdo}, but significant more work needs to be done in preparation for next-generation gravitational wave experiments. Several next-generation numerical relativity code are currently in development, employing novel methods that will enable high accuracy and performance on the supercomputers that will be available in the next decade (e.g. \cite{Fambri:2018udk, Kidder:2016hev, Daszuta:2021ecf}), but none of them have yet matured to the point where they can calculate gravitational waves or electromagnetic signals from merging neutron-star binaries.

\section{Modeling high-precision gravitational-wave observations}\label{sec:highprecision}
The next generation of gravitational-wave detectors on Earth and in space will yield observations of coalescing binary black holes with signal-to-noise ratios in the thousands, enabling high-fidelity observation of the behavior of the curved spacetime near stellar-mass black-hole horizons, the most strongly curved spacetime known. Gravitational wave signals will be so plentiful they will sometimes overlap.

As they have for current observations~\cite{TheLIGOScientific:2017qsa, Abbott:2020uma, Shibata:2017xdx}, numerical-relativity simulations will play crucial roles in the detection and interpretation of gravitational waves from merging black holes and neutron stars. In particular, waveforms from these simulations have been used to construct and validate approximate, phenomenological models necessary for interpreting observations (since numerical relativity is too costly to produce every model waveform needed)~\cite{Hannam:2013oca, Bohe:2016gbl, Khan:2015jqa, Blackman:2017pcm, Husa:2015iqa, Abbott:2020uma}, have featured in direct analysis of observations~\cite{Lange:2017wki}, and have helped validate our methods for detecting faint gravitational waves in detector data~\cite{Schmidt:2017btt}.

\begin{figure}
	\centering
	\includegraphics[width=5in]{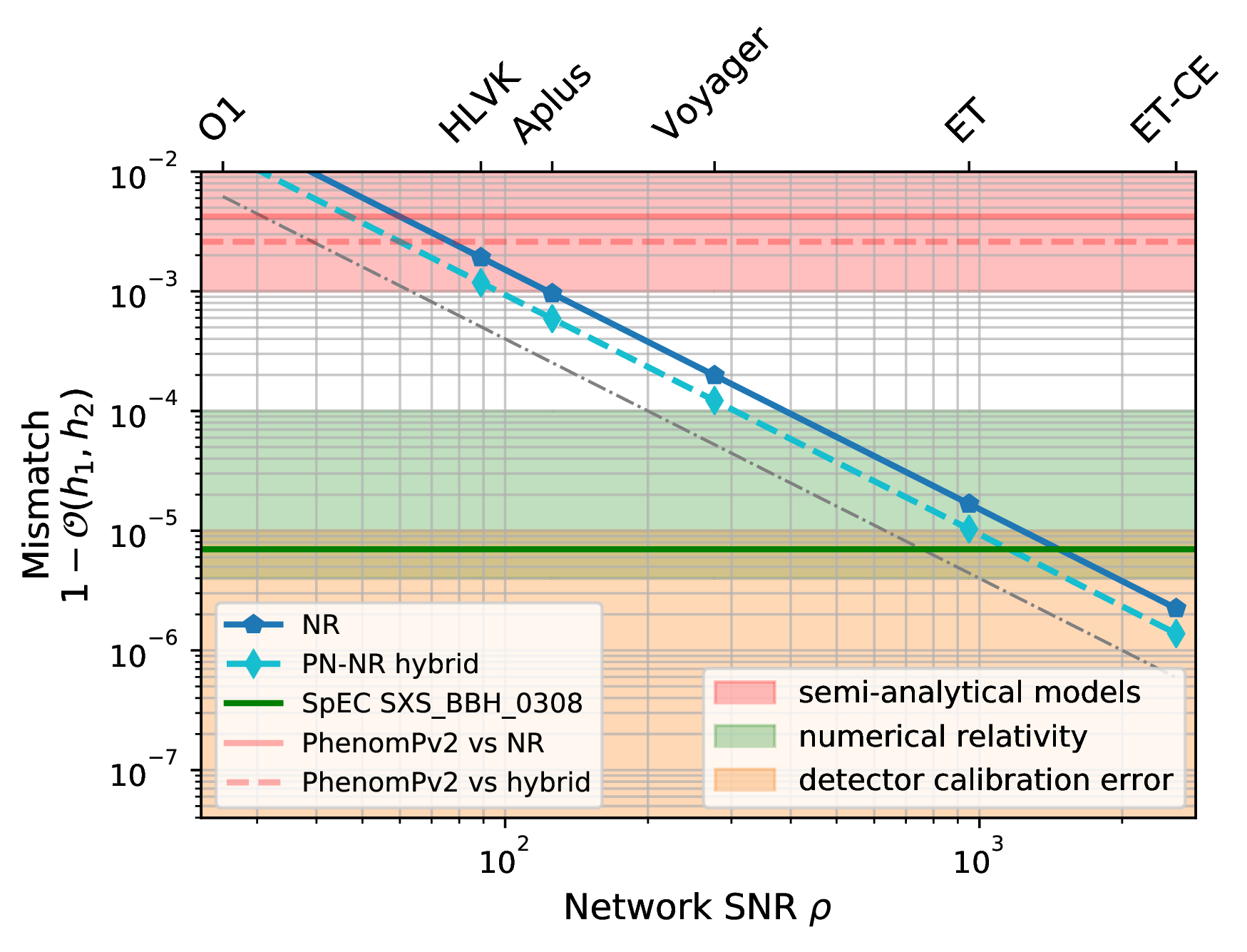}
\caption{\label{fig:AccuracyRequirements}
Predicted waveform accuracy for current second generation and future third generation detectors.
The mismatch between the waveform models is shown as a function of the detector signal-to-noise ratio (SNR).
Solid lines indicate results for pure numerical-relativity simulated signals, while dashed lines come from numerical-relativity signals extended (``hybridized'') with post-Newtonian (PN) waveforms in the inspiral. The blue lines and data points show how the mismatch falls with rising SNR.
Horizontal red lines show the mismatch of the signal against the IMRPhenomPv2 phenomenological template waveform at the signal parameters for LIGO's design sensitivity.
Taken from Fig.~2 of Ref.~\cite{Purrer:2019jcp}.
}
\end{figure}

But to model high-precision observations, numerical-relativity calculations will
have to be significantly more accurate than today's state of the art.
Qualitatively, the increase in accuracy is necessary to ensure that numerical
errors are smaller than experimental uncertainty given the much lower noise
level that next-generation gravitational-wave detectors will achieve (cf. Fig.~\ref{fig:AccuracyRequirements}).
Section 4
of Ref.~\cite{Boyle:2019kee} gives a quantitative estimate of how much the
accuracy must improve in terms of the improvement in signal to noise ratio,
based on a sufficient condition~\cite{Flanagan:1997kp, Lindblom:2008cm,
	McWilliams:2010eq, Chatziioannou:2017tdw} for a model waveform and an observed
gravitational waveform to be indistinguishable. Specifically, two gravitational
waveforms are indistinguishable if their mismatch $\mathcal{M}$ (a
noise-weighted inner product, defined, e.g., by Eq.~(24) of
Ref.~\cite{Boyle:2019kee}) is no larger than an amount proportional to the
inverse square of the signal-to-noise ratio $\rho$: $\mathcal{M} <
	D/\left(2\rho^2\right)$, where $D$ is a constant that depends on the number of
parameters needed to specify the gravitational waveform (e.g., in
Ref.~\cite{Boyle:2019kee}, for binary-black-hole waveforms, $D=8$). Thus
the accuracy required scales as the square of the signal-to-noise ratio $\rho$.

With today's detectors, the loudest gravitational waves from binary black holes
have $\rho \sim 24$, whereas future detectors will observe binary black holes
with $\rho$ in the thousands. The estimate in the previous paragraph suggests
that future detectors will demand more than an order of magnitude more accuracy
from numerical-relativity codes, even for modeling binary black holes
(a less challenging case than the case of simulations involving dense matter,
cf. Sec.~\ref{sec:ns}). Studies using
more sophisticated variants of the estimate sketched in the previous
paragraph~\cite{Purrer:2019jcp,Ferguson:2020xnm} give comparable
conclusions: numerical-relativity waveforms will need to be significantly
more sensitive to avoid introducing bias into interpretation of
high-precision gravitational-wave observations.

\section{Testing gravity in the nonlinear regime}\label{sec:testinggr}
    A consistent theory of quantum gravity is a major goal of modern physics. General relativity (GR) itself is not consistent with quantum mechanics,
because it breaks down at high-energy scales: it is non-renormalizable and exhibits physical singularities, such as those inside black holes and at the big bang. Candidate quantum-gravity theories include well-motivated extensions of GR, typically involving additional fields, higher curvature corrections, or symmetry breaking~\cite{Berti:2015itd,Clifton:2011jh,Alexander:2009tp,Yunes:2013dva,Will:2014kxa}.

Studies focusing on the formation or evolution of single black holes were considered in
Lorentz-violating theories~\cite{Garfinkle:2007bk,Adam:2021vsk},
massive gravity~\cite{Kocic:2019zdy, Torsello:2019jdg, Kocic:2019gxl, Torsello:2019tgc, Torsello:2019wyp, Kocic:2020pnm},
quadratic gravity~\cite{Benkel:2016rlz,Silva:2017uqg,Doneva:2017bvd,Ripley:2019irj,Ripley:2020vpk,Doneva:2020nbb,Doneva:2021dcc,Kuan:2021lol,Doneva:2021tvn,R:2022cwe},
or higher curvature effective theories~\cite{Held:2021pht}.
These have shown that black holes may develop scalar hair during their collapse, e.g., if described in quadratic and higher derivative gravity.
In contrast, in scalar-tensor theories it is neutron stars that can develop a scalar hair, while black holes may remain the same as their general-relativistic counterparts.
Given the extended phase-space of allowed, possibly hairy solutions
one might expect new signatures during the inspiral, merger and ringdown such as additional (scalar) radiation channels, a phase-shift of the gravitational wave emission as compared to the GR signal or new nonlinear effects during the merger.

    The nonlinear regime of gravity that unfolds during the collision of compact objects
is a particularly promising target to probe for extensions of GR, both because new phenomena are expected to be most prominent in that case~\cite{Cardoso:2012qm} and because candidate theories can be confronted with gravitational-wave observations~\cite{Yunes:2013dva,Berti:2015itd,Yunes:2016jcc,Yagi:2016jml,Barack:2018yly}.
    However, current gravitational-wave based tests of gravity
have either been limited to the weak-field regime or to null-tests against GR, because complete inspiral-merger-ringdown waveform models that capture these truly nonlinear beyond-GR effects are lacking.
    Numerical relativity
    has produced first proof-of-principle simulations beyond GR
in scalar-tensor theories~\cite{Barausse:2012da,Shibata:2013pra,Healy:2011ef,Berti:2013gfa,Sagunski:2017nzb},
Einstein-Maxwell-Dilaton models~\cite{Hirschmann:2017psw,Liebling:2019kpp},
cubic Horndeski theories~\cite{Figueras:2021abd},
effect field theories for dark energy, namely k-essence~\cite{terHaar:2020xxb,Bezares:2021dma,Bezares:2021yek},
dynamical Chern-Simons gravity~\cite{Okounkova:2017yby,Okounkova:2019dfo,Okounkova:2019zjf}, or
scalar Gauss-Bonnet gravity~\cite{Witek:2018dmd,Okounkova:2020rqw,Silva:2020omi,
East:2020hgw,East:2021bqk}.
A second body of work has studied the nonlinear dynamics of black-hole mimickers such as boson stars~\cite{Liebling:2012fv,Bezares:2017mzk,Palenzuela:2017kcg,Clough:2018exo,Dietrich:2018jov,Widdicombe:2019woy,DiGiovanni:2021ejn,Jaramillo:2022zwg,Bezares:2022obu,Bezares:2018qwa,Helfer:2018vtq}.
The effect of fluctuations near black holes' horizons,
mimicking for example microstate geometries, was modelled in Ref.~\cite{Liebling:2017pqs}.

An important difficulty when attempting simulations of binary mergers in beyond-GR theories is to devise mathematically well-posed and numerically stable formulations of the evolution equations. The development of formulations of Einstein's equations amenable to numerical simulations took decades to come to fruition~\cite{Shibata1995,Baumgarte99,Pretorius2005a,Lindblom:2007,Sarbach:2012pr,Hilditch:2013sba}, and repeating that work for every possible theory of gravity beyond general relativity is a daunting task.
For Brans-Dicke type scalar tensor theories it was proven that the resulting time evolution equations are indeed well-posed~\cite{Salgado:2008xh}.
More general scalar tensor theories of the Horndeski class can be cast in well-posed form if they are complemented with a modified generalized harmonic gauge as long as coupling parameters remain small~\cite{Kovacs:2020ywu}.
Other theories for which hyperbolic formulations are available include
$f(R)$ gravity~\cite{Mongwane:2016qtz},
or
Einstein-Aether theory~\cite{Sarbach:2019yso}.
On the other hand, there are a number of gravity theories that involve higher derivative terms that lead to (Ostrogradski) ghost instabilities and ill-posed evolution equations if they are treated as a complete theory.
For example, this has been shown
for dynamical Chern-Simons gravity, and one ``cure'' is to treat it as an effective field theory~\cite{Delsate:2014hba}.
Another remedy, proposed in Refs.~\cite{Cayuso:2017iqc,Allwright:2018rut},
and tested for a sixth-order model in~\cite{Cayuso:2020lca},
is a reformulation of the evolution equations in the spirit of Israel-Stewart theory for hydrodynamics.

Therefore, existing beyond-GR simulations have so far mainly focused on theories that can be recast as the evolution of a scalar or vector field coupled to the usual equations of general relativity (scalar-tensor, Maxwell-dilaton, boson stars), or on treating beyond-GR effects perturbatively~\cite{Okounkova:2017yby,Witek:2018dmd,Okounkova:2019zjf,Okounkova:2020rqw,Silva:2020omi}.
That said, the calculations in the decoupling limit have already identified new dynamical effects that have been missed with weak-field approximations.
This includes burst of scalar radiation during the merger of black holes in dynamical Chern-Simons gravity~\cite{Okounkova:2017yby}
or dynamical scalarization and descalarization of black holes in scalar Gauss-Bonnet gravity~\cite{Silva:2020omi}.
The gravitational waveform typically exhibits a phase-shift, compared to the vacuum GR case, due to additional radiation channels~\cite{Witek:2018dmd,Okounkova:2020rqw,East:2021bqk}.

Enabling high precision tests of gravity and searches for signatures of new physics will likely require innovative theoretical avenues to devise well-posed formulations of beyond-GR theories, and their application to creating high-precision catalogs of simulated waveforms.

\section{Black holes as cosmic particle detectors}\label{sec:particle}
    Although dark matter makes up more than 80\% of all matter in the universe,
    its nature,  composition and properties have remained elusive.
Black holes might shed light on the dark matter question and also ultralight beyond-standard model particles in general. Massive bosonic fields scattering off rotating black holes might form condensates around them if the fields' Compton wavelength is comparable to the black holes' size~\cite{Dolan:2007mj,Arvanitaki:2010sy,Brito:2015oca}. That is, astrophysical black holes in the mass range $5M_{\odot}\ldots10^{10}M_{\odot}$ are sensitive to ultralight particles in the mass range $10^{-21}{\rm{eV}}\ldots10^{-8}{\rm{eV}}$~\cite{Dolan:2007mj,Witek:2012tr,Arvanitaki:2010sy}. This range includes popular dark matter candidates~\cite{Hui:2016ltb},
    the QCD axion~\cite{Peccei:1977hh}
    and axion-like particles of the string axiverse~\cite{Arvanitaki:2009fg},
    as well as higher-spin fields such as
vector fields~\cite{Rosa:2011my,Witek:2012tr,Zilhao:2015tya,East:2017mrj,East:2017ovw,East:2018glu,Wang:2022hra}
or massive spin-2 fields~\cite{Brito:2020lup}; see also the companion Snowmass White Paper~\cite{Berti:2022wzk}.

    Because the underlying phenomenon of black hole superradiance only relies on gravitational interactions, it facilitates searches for new particles independently from their specific coupling to the standard model and thus complements traditional collider physics or direct detection experiments.
    The single black hole scenario has been studied extensively,
    and there are first computations of binary black-hole systems in the weak-field regime~\cite{Berti:2019wnn,Zhang:2019eid,Traykova:2021dua}, for extreme mass ratio inspirals~\cite{Baumann:2018vus,Baumann:2019ztm,Hannuksela:2018izj}
and in the fully nonlinear regime modeling the last orbit before merger, the merger and ringdown~\cite{Choudhary:2020pxy}.
    How these light fields impact the nonlinear dynamics of the late inspiral and coalescence of black-hole binaries endowed with scalar condensates and what its observational signatures are remain open questions. Addressing them will enable gravitational-wave based searches for new particles but will require significant advances in numerical relativity.


\section{Summary and future directions}\label{sec:conclusion} The next
generation of gravitational-wave detectors will probe fundamental physics with
exquisite sensitivity. Observations with far higher signal-to-noise ratios than
the loudest gravitational waves observed to date will use neutron-star mergers
to probe the nuclear physics of dense matter, use the loudest observations of
binary black holes to seek physics beyond general relativity, and will perhaps
enable a search for new particles that complements existing experimental
searches.

Realizing these goals will require model waveforms that will rely on a new
generation of numerical-relativity codes capable of achieving dramatically
improved accuracy. These codes will need to use novel techniques (such
as task-based parallelism) that enable
them to scale to make effective use of the exascale computing resources
expected to be available in the coming decade. Active development
of such codes is already underway. Examples of next-generation
numerical-relativity codes include NMesh~\cite{tichy2020numerical},
Dendro-GR~\cite{Fernando:2018mov}, GR-Athena++~\cite{Daszuta:2021ecf},
bamps~\cite{Bugner:2015gqa},
GRChombo~\cite{Clough:2015sqa,GRChombo}
and SpECTRE~\cite{deppe_nils_2021_4913286}.

Future studies will need to determine (more precisely than estimates such as in
Refs.~\cite{Purrer:2019jcp, Ferguson:2020xnm}) how the challenges of extremely
high signal-to-noise ratios and overlapping signals will impact the accuracy required to prevent
numerical-relativity simulations from biasing the interpretation of
next-generation gravitational-wave observations. One approach to such a study
would be to use numerical-relativity simulations to create simulated
gravitational-wave detections and then checking how much inaccuracies in model
waveforms used to interpret those signals bias the inferred properties.

These calculations will require significant computational resources to complete.
A typical numerical-relativity model waveform today typically require weeks to
months of runtime on tens to thousands of compute cores. Future waveforms will
require additional computational cost, in part because of higher accuracy
requirements (which will require higher resolution) and in part because future
detectors will have more sensitivity at lower frequencies, so that simulations
will have to be much longer to span the detectors' sensitive frequency spaces.
And many simulations will be necessary to span the parameter space of potential
signals. Binary-black-hole waveforms, for instance, are characterized by at least 7
parameters (the mass ratio and the black-hole spin angular momenta); even
spanning this space requires thousands of simulations (for instance, choosing 3
distinct possible values for each parameter would yield $3^7\approx 2,000$
simulations). Simulations involving neutron stars depend on even more
parameters, including the parameters characterizing the (not yet well
understood) neutron-star matter's equation of state. Simulations in theories
beyond general relativity also introduce additional parameters.

By meeting the challenges ahead, numerical relativity will play a crucial role in realizing the science goals of future gravitational-wave observatories, by enabling accurate, unbiased interpretations of their high-fidelity observations.

\section{List of Endorsers}
\input{Endorsers.tex}

	\bibliographystyle{unsrt}
	\bibliography{References}

\end{document}

%% file: Endorsers.tex
Cosimo Bambi (Fudan University) [bambi@fudan.edu.cn] \\
Sambaran Banerjee (University of Bonn) [sambaran@astro.uni-bonn.de] \\
Enrico Barausse (SISSA and INFN Sezione di Trieste, Italy) [barausse@sissa.it] \\
Wayne Barkhouse (University of North Dakota) [wayne.barkhouse@und.edu] \\
Daniele Bertacca (University of Padova and INFN Sezione di Padova, Italy) [daniele.bertacca@pd.infn.it] \\
Emanuele Berti (Johns Hopkins University) [berti@jhu.edu]\\
Miguel Bezares (SISSA and INFN Sezione di Trieste, Italy) [mbezares@sissa.it]\\
Alexander Bonilla (Universidade Federal de Juiz de Fora) [abonilla@fisica.ufjf.b] \\
Richard Brito (CENTRA, Instituto Superior T\'{e}cnico, Portugal) [richard.brito@tecnico.ulisboa.pt] \\
Liam Brodie (Washington University in Saint Louis) [b.liam@wustl.edu]\\
Marco Bruni (University of Portsmouth) [marco.bruni@port.ac.uk] \\
Tomasz Bulik (University of Warsaw) [tb@astrouw.edu.pl] \\
Manuela Campanelli (Rochester Institute of Technology) [manuela@astro.rit.edu]\\
Zhoujian Cao (Beijing Normal University) [zjcao@bnu.edu.cn] \\
Pedro R. Capelo (University of Zurich) [pcapelo@physik.uzh.ch] \\
Marco Cavagli\`a (Missouri University of Science and Technology) [cavagliam@mst.edu] \\
Jose A. R. Cembranos (Universidad Complutense de Madrid and IPARCOS, Spain) [cembra@ucm.es] \\
Philip Chang (University of Wisconsin-Milwaukee) [chang65@uwm.edu] \\
Maria Chernyakova (Dublin City University) [masha.chernyakova@dcu.ie]\\
Cecilia Chirenti (University of Maryland) [chirenti@umd.edu]\\
Katy Clough (Queen Mary, University of London) [k.clough@qmul.ac.uk] \\
Lucas G. Collodel (University of T\"ubingen) [lucas.gardai-collodel@uni-tuebingen.de] \\
Mesut \c{C}al{\i}\c{s}kan (Johns Hopkins University) [caliskan@jhu.edu] \\
Saurya Das (University of Lethbridge) [saurya.das@uleth.ca] \\
Tim Dietrich (University of Potsdam) [tim.dietrich@uni-potsdam.de] \\
Daniela Doneva (University of T\"ubingen) [daniela.doneva@uni-tuebingen.de] \\
Francisco Duque (CENTRA, IST, University of Lisbon) [francisco.duque@tecnico.ulisboa.pt] \\
Scott Field (University of Massachusetts Dartmouth) [sfield@umassd.edu] \\
Robert Eisenstein (Massachusetts Institute of Technology) [reisenst@mit.edu]\\
Zachariah B.~Etienne (University of Idaho) [zetienne@uidaho.edu] \\
Pedro G. Ferreira (University of Oxford) [pedro.ferreira@physics.ox.ac.uk] \\
Pau Figueras (Queen Mary University of London) [p.figueras@qmul.ac.uk] \\
Giacomo Fragione (Northwestern University) [giacomo.fragione@northwestern.edu]\\
J{\"o}rg Frauendiener (University of Otago) [joerg.frauendiener@otago.ac.nz] \\
Mandeep S. S. Gill (Stanford University) [msgill@slac.stanford.edu] \\
Andreja Gomboc (University of Nova Gorica) [andreja.gomboc@ung.si] \\
Leonardo Gualtieri (University of Rome ``La Sapienza'') [leonardo.gualtieri@roma1.infn.it]\\
Roland Haas (University of Illinois) [rhaas@illinois.edu] \\
Alexander Haber (Washington University in Saint Louis) [ahaber@physics.wustl.edu]\\
Wen-Biao Han (Shanghai Astronomical Observatory, CAS) [wbhan@shao.ac.cn] \\
Mark Hannam (Cardiff University) [hannammd@cardiff.ac.uk] \\
Ian Hawke (University of Southampton) [I.Hawke@soton.ac.uk] \\
Lavinia Heisenberg (Heidelberg University/ETH Zurich)[laviniah@ethz.ch] \\
Thomas Helfer (Johns Hopkins University) [thelfer1@jhu.edu] \\
Shang-Jie Jin (Northeastern University, China) [jinshangjie@stumail.neu.edu.cn]\\
Cristian Joana (University of Louvain) [cristian.joana@uclouvain.be] \\
Atul Kedia (Rochester Institute of Technology) [asksma@rit.edu] \\
Antoine Klein (University of Birmingham) [antoine@star.sr.bham.ac.uk] \\
Shiho Kobayashi (Liverpool John Moores University) [s.kobayashi@ljmu.ac.uk]\\
Kostas Kokkotas (University of T\"ubingen)[kostas.kokkotas@uni-tuebingen.de] \\
Savvas M. Koushiappas (Brown University) [koushiappas@brown.edu] \\
Dicong Liang (Peking University) [dcliang@pku.edu.cn] \\
Steven L. Liebling (Long Island University) [steve.liebling@liu.edu] \\
Eugene A. Lim (King's College London) [eugene.a.lim@gmail.com] \\
Hyun Lim (Los Alamos National Lab) [hyunlim@lanl.gov] \\
Nicole Lloyd-Ronning (Los Alamos National Lab) [lloyd-ronning@lanl.gov] \\
Zhengwen Liu (DESY Hamburg) [zhengwen.liu@desy.de] \\
Carlos O. Lousto (Rochester Institute of Technology) [colsma@rit.edu] \\
Ilya Mandel (Monash University) [ilya.mandel@monash.edu]\\
Irvin Martínez (University of Cape Town)[mrtirv001@myuct.ac.za]\\
Andrea Maselli (Gran Sasso Science Institute) [andrea.maselli@gssi.it] \\
Cole Miller (University of Maryland) [miller@astro.umd.edu]\\
Elias R. Most (Princeton University) [emost@princeton.edu] \\
Bishop Mongwane (University of Cape Town) [bishop.mongwane@uct.ac.za] \\
Suvodip Mukherjee (Perimeter Institute) [smukherjee1@perimeterinstitute.ca]\\
Hiroyuki Nakano (Ryukoku University)
[hinakano@law.ryukoku.ac.jp] \\
David Neilsen (Brigham Young University) [david.neilsen@byu.edu] \\
David A.\ Nichols (University of Virginia) [david.nichols@virginia.edu]\\
Rafael C. Nunes (Instituto Nacional de Pesquisas Espaciais, Brazil) [rafael.nunes@inpe.br]
\\Maria Okounkova (Flatiron Institute) [mokounkova@flatironinstitute.org]
\\Richard O'Shaughnessy (Rochester Institute of Technology) [rossma@rit.edu] \\
Giorgio Orlando (University of Groningen) [g.orlando@rug.nl] \\
N\'estor Ortiz (National Autonomous University of Mexico) [nestor.ortiz@nucleares.unam.mx] \\
Benjamin J. Owen (Texas Tech University) [benjamin.j.owen@ttu.edu] \\
Carlos Palenzuela (Universitat de les Illes Balears) [carlos.palenzuela@uib.es] \\
Harald P. Pfeiffer (Max Planck Institute for Gravitational Physics (Albert Einstein Institute)) [harald.pfeiffer@aei.mpg.de] \\
Denis Pollney (Rhodes University) [d.pollney@ru.ac.za] \\
Geraint Pratten (University of Birmingham) [g.pratten@bham.ac.uk] \\
Alireza Rashti (Florida Atlantic University) [arashti2016@fau.edu] \\
Luciano Rezzolla (Goethe University, Frankfurt am Main, Germany)[rezzolla@itp.uni-frankfurt.de] \\
Angelo Ricciardone (University of Padova)[angelo.ricciardone@pd.infn.it] \\
Justin L. Ripley (University of Cambridge) [jr860@cam.ac.uk] \\
Dorota Rosinska (University of Warsaw)[drosinska@astrouw.edu.pl]\\
Stephan Rosswog (Stockholm University) [stephan.rosswog@astro.su.se] \\
Milton Ruiz (University of Illinois at Urbana-Champaign) [ruizm@illinois.edu] \\
Violetta Sagun (University of Coimbra) [violetta.sagun@uc.pt]\\
Mairi Sakellariadou (King's College London) [mairi.sakellariadou@kcl.ac.uk] \\
Zeyd Sam (University of Potsdam) [zeyd.sam@uni-potsdam.de] \\
Nicolas Sanchis-Gual (University of Valencia) [nicolas.sanchis@uv.es] \\
Misao Sasaki (University of Tokyo) [misao.sasaki@ipmu.jp]\\
B.S. Sathyaprakash (Penn State) [bss25@psu.edu] \\
Erik Schnetter (Perimeter Institute for Theoretical Physics) [eschnetter@perimeterinstitute.ca] \\
Lijing Shao (Peking University) [lshao@pku.edu.cn] \\
Stuart L. Shapiro (University of Illinois) [slshapir@illinois.edu] \\
Deirdre Shoemaker (University of Texas at Austin) [deirdre.shoemaker@austin.utexas.edu] \\
Daniel M. Siegel (Perimeter Institute for Theoretical Physics) [dsiegel@perimeterinstitute.ca] \\
Hector O. Silva (Max Planck Institute for Gravitational Physics (Albert Einstein Institute)) [hector.silva@aei.mpg.de] \\
Joshua Smith (California State University, Fullerton) [josmith@fullerton.edu] \\
Carlos F. Sopuerta (Institute of Space Sciences, ICE-CSIC and IEEC) [carlos.f.sopuerta@csic.es] \\
Ulrich Sperhake (University of Cambridge)
[U.Sperhake@damtp.cam.ac.uk]\\
Leo Stein (Univ. of Mississippi) [lcstein@olemiss.edu] \\
Nikolaos Stergioulas (Aristotle University of Thessaloniki) [niksterg@auth.gr] \\
Ling Sun (The Australian National University) [ling.sun@anu.edu.au] \\
Gianmassimo Tasinato (Swansea University) [g.tasinato@swansea.ac.uk] \\
Maximiliano Ujevic Tonino (Universidade Federal do ABC) [mujevic@ufabc.edu.br] \\
Elias C. Vagenas (Kuwait University) [elias.vagenas@ku.edu.kw]\\
Alex Vano-Vinuales (CENTRA, IST, University of Lisbon) [alex.vano.vinuales@tecnico.ulisboa.pt] \\
Daniele Vernieri (University of Naples and INFN Sezione di Napoli, Italy) [daniele.vernieri@unina.it] \\
Flavio Vetrano (University of Urbino , Italy) [flavio.vetrano@uniurb.it] \\
Anzhong Wang (Baylor University, Waco, Texas) [anzhong\_wang@baylor.edu] \\
Barry Wardell (University College Dublin) [barry.wardell@ucd.ie] \\
Scott Watson (Syracuse University) [gswatson@syr.edu] \\
Vojt{\v e}ch Witzany (University College Dublin) [vojtech.witzany@ucd.ie] \\
Kinwah Wu (University College London) [kinwah.wu@ucl.ac.uk] \\
Shu-Cheng Yang (Shanghai Astronomical Observatory, CAS) [ysc@shao.ac.cn] \\
Stoytcho Yazadjiev (University of Sofia) [yazad@phys.uni-sofia.bg] \\
Anzhong Wang (Baylor University) [anzhong\_wang@baylor.edu]\\
Alan J.~Weinstein (Caltech) [ajw@caltech.edu] \\
Ivonne Zavala (Swansea Unievrsity) [e.i.zavalacarrasco@swansea.ac.uk]\\
Miguel Zilhão (Universidade de Aveiro) [mzilhao@ua.pt] \\
Junjie Zhao (Beijing Normal University) [junjiezhao@bnu.edu.cn] \\